# The myth of the Digital Earth
# between fragmentation and wholeness

Andrea Ballatore
Center for Spatial Studies
University of California, Santa Barbara
aballatore@spatial.ucsb.edu

**Abstract**

Daring predictions of the proximate future can establish shared discursive frameworks, mobilise capital, and steer complex processes. Among the prophetic visions that encouraged and accompanied the development of new communication technologies was the "Digital Earth," described in a 1998 speech by Al Gore as a high-resolution representation of the planet to share and analyse detailed information about its state. This article traces a genealogy of the Digital Earth as a techno-scientific myth, locating it in a constellation of media futures, arguing that a common subtext of these envisionments consists of a dream of wholeness, an afflatus to overcome perceived fragmentation among humans, and between humans and the Earth.

## 1. Introduction

The ability to collect and communicate information about the Earth is an essential element to human material survival. Hence, it should not come as a surprise that the development of digital media over recent decades has impacted enormously on how Earth-related information is generated, analysed, and stored. Positioning infrastructures, remote sensor networks, geographic information systems, Web-enabled smartphones, virtual globes, digital maps, SatNavs, and location-based services form complex technological assemblages that impact human mobility, reshaping travel modes, geographical imaginations, cultural flows, and the intimate connections between places and people. To capture these developments, human geographers have named these recent trends as "new spatial media", "neogeography", and the "geoweb" (Leszczynski and Wilson 2013), while communication scholars recently coined the term "Earth-observing media" (Russill 2013). Such technologies do not exist in a coherent framework and did not develop from a single research programme, but rather have arisen from diverse military, corporate, and academic milieux, for often radically different purposes.[1]

Describing detailed visions of the future might be thought of as an activity of a pre-modern past, incompatible with modern techno-scientific enterprises. Nothing could be farther from the truth. From the perspective of media studies, it is particularly important to look at what visions animated the development and the adoption of geo-technologies, uncovering the specific techno-scientific and political discourses that surrounded their inception. As Dourish and Bell (2011) noted, research on new information technology is often animated by "organising visions" on how the new tools will impact the world, bringing about exciting and inevitable changes not only for technologists, but for all humankind (p. 1). These pictures of the "proximate future" tend to be portrayed as saturated by new technology, always around the corner, therefore making the present outdated (Bell and Dourish 2007). Analysing visions of future mobile communication, de Vries (2012) defines them as "necessary fictions," resulting from the "psychological need to create order out of chaos, to create utopian landmarks that we can look out for while travelling along the paths of life" (p. 48).

A long-standing tradition in media scholarship has focused on this recurrence of utopian (and dystopian) visions around virtually every major technical invention, including electricity, wireless telegraphy, steam power, air travel, telephony, cinema, television, nuclear energy, and space travel (e.g. Marvin 1990, Dourish and Bell 2011). Each time, the new tools promised to eradicate major societal ills, such as war, poverty, isolation, scarcity of resources, and gender, racial, or economic inequalities. A key concept within this tradition is the notion of the "technological sublime," (Marx 1964, Nye 1996), i.e. the sentiment of wonder triggered by the contemplation of technology and mechanisation as a specific character of American culture, fostering faith in progress and modernity. Recasting the concept in the context of electricity as the "electrical sublime," Carey and Quirk (1989a) observed that electricity promised "the same freedom, decentralization, ecological harmony, and democratic community that had hitherto been guaranteed but left undelivered by mechanization" (p. 94).

The emergence of digital computers in the mid-20th century is no exception to the rule, and digital media have fostered wild narratives of the future, particularly in relation to the development of the Internet. Turner (2006) identified the common origin of these narratives as "digital utopianism," and traced it back to countercultural movements in the 1960s. These hyped digital narratives are analysed by Coyne (1999) under the lens of "technoromanticism," that is the "spectrum of romantic narrative that pervades the digital age," inflating expectations and promoting the heroism of the digital entrepreneur (p. ix). Mosco (2004) strikingly explored computer-centred utopian visions under the notion of the "digital sublime," reading these technological discourses as modern myths, drawing on anthropological theories of myth-making as social construction of meaning. Myths are animated by a "bricoleur," an individual who "pulls together the bits and pieces of technology's narratives, to fashion a mobilizing story for our time … [an] heroic narrative with didactic effect" (p. 36). In this view, whether specific visions would be fulfilled or not is beyond the point:

> it is important to state at the outset that myths mean more than falsehoods or cons; indeed, they matter greatly. Myths are stories that animate individuals and societies by providing paths to transcendence that lift people out of the banality of everyday life. They offer an entrance to another reality, a reality once characterized by the promise of the sublime. … myths are not true or false, but living or dead (Mosco 2004, p. 3).

Following the tenets of this strand of media scholarship, this article offers an analysis of the Digital Earth, a technological vision that influences the development of prominent geo-technologies, evoked by Al Gore in the late 1990s. The next section describes the origin and the tenets of the Digital Earth vision in a mythical framework, and the context in which Al Gore conceived it. Section 3 compares the Digital Earth with other influential visions of Earth, including McLuhan's Global Village and Lovelock's Gaia hypothesis, arguing that the underlying drive of these visions is a dream of wholeness. The utopian subtext in the Digital Earth vision is framed in a constellation of media futures in Section 3. Section 4 identifies recurring dreams of wholeness that pervade envisionments of new media, whose limitations are addressed in Section 5. Finally, Section 6 provides a closure to this discussion.

## 2. Al Gore, myth-maker

As Dourish and Bell (2011) pointed out, the visions of the future are often more interesting for what they say about the present rather than about the future. When discussing the development of digital media in the late 20th century, the role of Albert Gore, Jr. can hardly be underestimated. In his political career, Gore was elected several times to the U.S. House of Representatives from 1976 with the Democratic party, as a senator since 1984, and was appointed Vice President in 1993, serving two full mandates with Bill Clinton until 2001, acting as his chief advisor. Under the influence of his father, who in the 1950s mixed a concern for soil erosion with an interest for nuclear technology, Gore constructed his politics by combining environmentalism and new technologies (Turque 2000). The period of his vice-presidency (1993-2001), in which Al Gore promoted the Digital Earth and other new media-fuelled visions, was ripe for technological narratives of progress. As Barbrook (2007, p. 268-70) argued, the Clinton administration developed the view that the US would spread political consensus, multicultural tolerance, and market economics across the globe. The "systemic universalization of digital network capitalism" was an essential ingredient to bring about the end of history (Franke 2013, p. 14). The optimism of the post-Cold War neoliberal order was epitomised by the 1997 *Wired* cover, stating "We're facing 25 years of prosperity, freedom, and a better environment for the whole world. You got a problem with that?"

## 2.1. Championing the Information Age

Although Gore did not "invent the Internet" as he is accused of claiming, he acted as a "fierce champion of federal support of computer networking," tirelessly advocating the concept of an "Information Superhighway," a metaphor for the Internet as a network of interconnected networks, echoing the federal programme to build the American highway network in the 1950s (Ceruzzi 2003, pp. 321-3). The establishment of a digital high-speed network had had high priority since the early 1990s, with an explicit call for federal funding to create an infrastructure to reach McLuhan's Global Village (Gore 1991). In 1994, Gore introduced his vision for the Global Information Infrastructure (GII) at the first World Telecommunication Development Conference held in Buenos Aires, arguing for each nation to develop the GII with private investment, competition, light regulation, and universal access (Gore 1994). In this speech, he claimed that for 150 years people have tried "to wrap nerves of communications around the globe, linking all human knowledge" with communication technologies, and stated that finally the time had come for this vision to come true, bringing "all the communities of the world together" towards a "new Athenian Age of democracy" (p. na).

Through his advocacy of the Information Superhighway, Gore became, as Mosco (2004) put it, one of "the top bricoleurs of cyberspace" (p. 38). In office, he acted as a new media entrepreneur and corporate advisor, constantly crossing the boundaries between politics, academia, and high-tech corporations. Thanks to his direct knowledge of these worlds, Gore's technological visions aptly combined the utopian hope in technological progress as a vehicle of modernisation, pervaded by the digital sublime, with pragmatic free-market economic principles to fill the new media platforms with capital, reaching diverse audiences with seemingly neutral, desirable, and uncontroversial goals for the proximate future.[2] While media scholars have extensively discussed the Information Superhighway, little attention has been devoted to another vision animated by Gore in the same period, that of the Digital Earth.

## 2.2. The Digital Earth vision

Al Gore developed his interest in environmentalism, ecology, and climate change early on in his political career in the 1970s, and actively tackled a number of environmental issues in his constituency. Gore's environmentalism is firmly inscribed in a liberal-capitalist worldview, advocating market-based solutions to the climate crisis (Ross 1994). As Vice President, his efforts to build the Information Superhighway merged with his environmental vision of an environmentally sustainable capitalism. Combining new digital media and his concern for the state of the Earth, he first described the idea of a "Digital Earth" in his 1992 best-selling book, *Earth in the Balance* (Gore 1992). To tackle the impending environmental catastrophe, Gore argued, a new global Marshall Plan should be initiated by the U.S., combining economic growth with ecological responsibility, and renewing the United States' dedication to social justice, democracy, and free-market economics towards an "environmentalism of the spirit" (Luke 1998).

To understand what was happening to the planet, Gore outlined a "Digital Earth program," aimed at building a "new global climate model," drawing on heterogeneous data sources (Gore 1992, p. 358). Using distributed parallel computation, enormous quantities of data could be used to predict the Earth's climate with unprecedented precision. He later developed this idea and described it in detail in the 1998 speech at the California Science Center in Los Angeles (Gore 1998). To make sense of the "flood of geospatial information" produced every day, there was a need for a "multi-resolution, three-dimensional representation of the planet, into which we can embed vast quantities of geo-referenced data" (p. na). Gore illustrated the vision with a rhetorical move towards the digital sublime:

> Imagine, for example, a young child going to a Digital Earth exhibit at a local museum. After donning a head-mounted display, she sees Earth as it appears from space. Using a data glove, she zooms in, using higher and higher levels of resolution, to see continents, then regions, countries, cities, and finally individual houses, trees, and other natural and man-made objects. Having found an area of the planet she is interested in exploring, she takes the equivalent of a "magic carpet ride" through a 3-D visualization of the terrain. ... She can get more information on many of the objects she sees by using her data glove to click on a hyperlink. To prepare for her family's vacation to Yellowstone National Park, for example, she plans the perfect hike to the geysers, bison, and bighorn sheep that she has just read about. (Gore 1998, p. na)

The young girl cannot only move through space, but can also travel in time, in a cross-medial dream. "After taking a virtual field-trip to Paris to visit the Louvre, she moves backward in time to learn about French history, perusing digitized maps overlaid on the surface of the Digital Earth, newsreel footage, oral history, newspapers and other primary sources" (p. na). This vision of a child delving into the digital sublime echoes early comments on the potential of television, which would "put children on a magic carpet," enabling them to fly to any part of the world, stimulating their scientific talent (Mosco 2004, p. 135). In order to make the Digital Earth come true, Gore continued, concerted efforts around several technologies were needed, including computers, mass storage, satellite imagery, and interoperable data formats. In this move, Gore acted as the bricoleur, combining existing pieces of loosely related technologies to conjure up a meaningful new future horizon.

The speech also illustrated the potential applications of the Digital Earth, outlining a range of promises, considerably consistent with the utopian visions fuelled by old media when they were new. Virtual diplomacy would be conducted on the Digital Earth, solving border disputes by visualising new political scenarios on virtual globes. Geo-localised crime information would help the police fight deviant behaviour. Abundant environmental information would assist administrations in preserving biodiversity. The Digital Earth would provide support in modelling and predicting climate change, and would increase agricultural productivity. The vision was concluded with typical Gore's combination of a somewhat naive image of a better society (the Digital Earth will enable "our children to learn more about the world around them") with liberal-capitalist pragmatism (the Digital Earth will "accelerate the growth of a multi-billion dollar industry").

## 2.3. Keeping the Digital Earth spinning

To fulfil his vision, Gore used his executive power as Vice President to launch the "Digital Earth Initiative" (DEI), in 1998, chaired by NASA and involving several U.S. federal agencies around issues of data standardisation (Foresman 2008). When he lost his bid for the office of president in 2000, the newly elected Bush government promptly cut all federal funding to the DEI by 2001. However, efforts towards the development of virtual globes continued in corporations and prototypes appeared in the early 2000s, including the CIA-funded Keyhole's *Earthviewer*, acquired by Google in 2005 and rebranded as *Google Earth*, NASA's *World Wind*, Microsoft's *Virtual Earth*, and ESRI's *ArcGIS Explorer*. These products, particularly Google Earth which reached widespread popularity, partly implement the Digital Earth vision, enabling a fast and interactive navigation of the planet. While Gore diverted his attention to other issues, others acted as bricoleurs to update and keep the vision of Digital Earth alive.

A group of leading geographers, Earth and information scientists, including Michael Goodchild, Lu Yongxiang, and Tim Foresman, launched a number of initiatives, including the International Journal of Digital Earth, the International Society for Digital Earth,[3] and a series of international Summits to build a community around this hopeful technological vision. In this context, several efforts have been made to revise and update the original Digital Earth vision based on recent changes in the technological landscape, re-defining techno-scientific agendas toward a "next generation" Digital Earth (Craglia et al. 2008, Goodchild et al. 2012, Craglia et al. 2012). Regardless of the actual delivery of its promises, the myth of the Digital Earth is most definitely still alive.

## 3. Digital Earth as a future utopia

To draw a genealogy of Digital Earth vision, it is necessary to inscribe it in the future-oriented images that have been triggered by new media technology. The purpose here is not to dismiss the Digital Earth vision as unworthy or impossible, but rather to frame it within a broader context of media discourses of great cultural significance. The spectacle offered by techno-futuristic visions bring the audiences to what Mosco (2004) calls a "liminal state" between the out-dated present and an exciting proximate future. In this sense, although the Digital Earth is less hyperbolic and implausible than analogous Silicon Valley visions, it nonetheless bears traces of a utopian hiatus. The invocation of images of the future can be interpreted by what Manuel and Manuel (1979) called the "utopian propensity," rejecting fixed definitions of the term, and arguing that it is difficult to distinguish between future utopias and "nostalgia for an idealized bygone human condition" (p. 5).

In the history of media, such "organising utopias" should not be read only negatively as chimerical and dangerously false fantasies, but rather as serious and instrumental future-oriented narratives that drive and mobilise resources, capital, and common citizens to develop and adopt new communication technologies. The utopian propensity, it can be argued, is a crucial element that drives techno-scientific agendas in politics, academia, and industry. As Borup et al. (2006) noted, scientists' expectations about the future contribute to drive their scientific activity, setting agendas, defining roles, clarifying duties, and creating new opportunities. To put the Digital Earth vision in context, it is useful to discuss it in relation to major media narratives that emerged in the 20th century.

### 3.1. A constellation of digital media futures

The Digital Earth is only a recent node added to a dense graph of future media narratives revolving around the planet. These envisionments read the relationship between media and the planet from two complementary perspectives. On the one hand, digital media are employed to build increasingly detailed models of the Earth and its processes, which is the core idea of the Digital Earth. In this sense, social media enable what Graham (2010) calls the construction of a "virtual Earth," mirroring and augmenting the geographic reality. On the other hand, the actual spread of communication networks - and the resulting space-time compression - fuels visions of the Earth *itself* becoming "alive" or "intelligent" thanks to its man-built nervous system.

Arguably the most enduring media vision of Earth, the metaphor of the "Global Village" was developed by Marshall McLuhan (1962) to describe the cultural changes brought about by electronic telecommunication networks. Although McLuhan did not see the Global Village as free of conflicts and crises, the metaphor was deployed by American neoliberal forces in the 1990s to promote the vision of a harmonious single post-Cold War market under a single form of government (Barbrook 2007). To justify the federal support for the technological development of the Internet, Al Gore (1991) called it the "infrastructure for the Global Village" (p. 150).

Foresman (2008) traced the inception of the Digital Earth back to Buckminster Fuller. In the 1960s, the eclectic architect conceived the "geoscope" as a supreme tool of global simulation, relying on a conceptualisation of the planet as a cybernetic system. Conceived as a computerised control room, the "geoscope" consisted of a geodesic miniature Earth, covered with aerial photographs of its surface and millions of little light bulbs, which would visualise an endless range of natural and human phenomena, promoting an ecological ethos:

> This 200-foot-size Geoscope would make it possible for humans to identify the true scale of themselves and their activities on this planet. Humans could thus comprehend much more readily that their personal survival problems related intimately to all humanity's survival. ... With the Geoscope humanity would be able to recognize formerly invisible patterns and thereby to forecast and plan in vastly greater magnitude than heretofore. ... The consequences of various world plans could be computed and projected, using the accumulated history-long inventory of economic, demographic, and sociological data. All the world would be dynamically viewable and picturable and radioable to all the world (Fuller 1981, pp. 161-97).

Fuller envisioned the Geoscope as the test bed of a World Game, in which teams would test their scientific theories with the simulation engine on an enormous database of natural and human data. The theory bringing world peace and prosperity would win, and enable politicians to take rational, unbiased decisions. This desire to visualise the entire planet is central to the emergence of what Turner (2006) calls "digital utopianism" from countercultures in California. One of the leaders of the counterculture, Steward Brand, prominently featured pictures of the Earth taken from space in his *Whole Earth Catalog* magazine in the the late 1960s, aiming at "a grand new synthesis of everything, yet particularly technology and nature" (Franke 2013, p. 15).

In the new consciousness of the counterculture, the NASA photographs of the entire Earth taken from space fuelled a sublime impression of interconnectedness and interdependence between humans and the planet, fostering cybernetic visions of ecological harmony (Franke 2013). The desire to reconcile human flourishing and nature was further fuelled by the Gaia hypothesis, the controversial view of Earth and its occupants as a homeostatic, living organism, advanced by scientist and environmentalist James Lovelock (Lovelock and Margulis 1974). Conceived indeed as a scientific theory, the Gaia hypothesis inspired a plethora of New Age environmentalist and spiritual ideas. As Kruger (2007) pointed out, Gaia intermingled with narratives of the Internet with religious undertones, in which digital media are seen as a nervous system extended across the planet, with the Earth becoming a giant and omniscient consciousness. This "globe clothing itself with a brain," as Kreisberg (1995) expressed it, has tenuous intellectual roots in Pierre Teilhard de Chardin's *noosphere*, and would enable a new holistic relationship with the Earth. Among many other theorists of Web-based "collective intelligence" (e.g. Lévy & Bonomo 1999), system theorist Francis Heylighen (2002) furthered this cyber-utopia into a full-fledged Global Brain, promising nothing less than a "much enhanced level of consciousness and a state of deep synergy or union that encompasses humanity as a whole" (p. 3). Although varying in scope and content, these visions share with the Digital Earth a deep re-conceptualisation of the planet as a digital whole.

## 4. Dreams of wholeness

Digital visions of the Earth differ greatly in utopian intensity, but are all concerned with overcoming a current state of perceived fragmentation towards *wholeness*. The recurrence and persistence of techno-scientific myths can be accounted for by observing that, in human communication, myths constitute effective rhetorical schemes to convey wholes, when particulars are fragmented and complex (Sykes 1970). Although concepts such as the Nye's technological sublime, Coyne's technoromanticism, and Mosco's digital sublime allude to the need for wholeness, they do not adequately analyse it. As Bell and Dourish (2007) put it, "homogeneity and an erasure of differentiation is a common feature of future envisionments" (p. 134), whilst Coyne (1999) remarked that digital narratives present multiplicity pejoratively as fragmentation. More broadly, Wood (2010) argues that post-war American culture has been characterised by holistic worldviews which "worked to unite what others separated and bring together what others treated separately: mind, body, and spirit; individuals and community; human beings and nature; nature and technology; science and religion; the material world and the sacred" (p. 4).

Extending his ontological holism to human affairs, theoretical physicist and philosopher David Bohm put wholeness and fragmentation at the core of his views. Whilst human beings have always been seeking physical, social, and psychological wholeness, states of fragmentation occur causing crises. Science keeps generating incompatible sub-disciplines, resulting in epistemological fragmentation (Bohm 1985). At the social level, human consciousness is split across nations, ethnic groups, classes, professions, and families, stressing a harmful view of a separate Ego that prevents cooperation (Bohm 1980). The natural environment is seen as an aggregate of separate parts to be used for different human purposes, enabling its large-scale overexploitation and destruction. Religions, beyond superficial doctrinal differences, promise wholeness, but keep breaking up into incoherent and hostile factions (Bohm 1985).

Overwhelmed by fragmentation, individuals are driven to helplessness, despair, and neurosis. Unsurprisingly, then, a strategy to cope with fragmentation consists of projecting myths of an earlier golden age, "before the split between man and nature and between man and man had yet taken place" (Bohm 1980, p. 3). The development of new digital media presents an opportunity to react against fragmentation, aiming at grasping the "implicate order" which lies behind a meaningless surface. Hence, bricoleurs animate technological myths by showing how new media will dissolve fragmentation into dreams of wholeness. New media-fuelled envisionments such as Al Gore's Digital Earth strive for wholeness along several dimensions:

**Social wholeness.** New media are seen as generating a planetary society without internal conflicts, based on a "communication utopia," in which there can be no misunderstanding (de Vries 2012). Political wholeness has been promised in many forms, for example as a "new Athenian democracy" by Al Gore (1994). New media technologies are also expected to solve societal ills by producing a "cornucopia of jobs, markets, and products, to rejuvenate ailing economies, to refund declining universities, to reemploy the unemployed and redundant" (Carey and Quirk 1989a, p. 89).

**Technological wholeness.** Real technology exists in unstable, chaotic, heterogeneous assemblages, fragmented by socio-economic and technical divides. Tools that are unimpressive in their fragmented isolation are combined into awe-inspiring wholes, which attract technologists whose daily work is plagued by interoperability issues. For example, through the Digital Earth, several existing technologies, scientific practices, and data sources could merge into a coherent framework, breaking up the "silos" (Foresman 2008).

**Epistemological wholeness.** Incompatible and competing theories are generated by scientific disciplines (Sarewitz 2010). As cybernetics has established a scientific framework to describe complex systems in terms of information processing, new media are seen as vehicles to promote shared epistemological frameworks, bringing scientists and scholars together. The Digital Earth vision, like Fuller's Geoscope, is seen as a platform for scientists to develop a common language to talk about the planet and its inhabitants. Total epistemological wholeness corresponds with omniscience, a religious afflatus present in many techno-futuristic envisionments.

**Ecological wholeness.** Reconciling humans with their environment in a Gaian homeostasis is a major discursive element in the Digital Earth. Unsurprisingly, environmentalism is deeply pervaded by utopian visions of wholeness (Pepper 2005).

While radical, "deep" environmentalism with anarchist, situationist, or neo-primitivist roots tends to be technologically pessimistic, liberal environmentalism such as that of Al Gore sees new media as an important vehicle to perfect capitalism and integrate it smoothly with the Earth. The Digital Earth does not bring ecological wholeness *by itself*, but it helps towards this objective, for example by identifying and effectively communicating the dangers of global warming.

**Spiritual wholeness.** Although rarely invoked explicitly, the discourse of spiritual wholeness pervades most envisionments of media futures (Kruger 2007). Digital media are seen as enablers of divine wholeness in a global religious community, responding to increased secularisation and loss of local communities. Religious feelings can also be mobilised towards ecological wholeness. A devout Baptist, Al Gore called for an "environmentalism of the spirit," advocating a panreligious, Earth-centered view (Gore 1992, pp. 258-9). Although the reflection of the God is difficult to see in the tiny corners of Creation, argued Gore, it is nonetheless present in its entirety, similarly to a vivid hologram composed of many small faint images.

## 5. The limits of wholeness

The dreams of future wholeness between humans and the Earth described so far are important, and yet entirely fictional, constructions. At the core of the utopian propensity lies a crucial paradox. Totalising solutions, as that which de Vries (2012, pp. 48-56) discusses at length, while promising to reconcile fragments into a homogenous whole, inevitably create new fragmentation. If techno-utopian envisionments present accomplished solutions to social conflict, epistemological fractures, and ecological catastrophes, they do not offer substantial mechanisms to reduce the complexity in the interplay between new technologies and shifting social forces. Fragmentation is, indeed, an inevitable condition in human affairs, and, as even Bohm (1980) admits, fragmentation is necessary "to reduce problems to manageable proportions" (p. 2). The cultural spaces opened by the invention of new technologies are *loci* suitable for imagining and longing for golden ages of wholeness.

Holistic worldviews have their merits and have indeed inspired more inclusive, progressive politics (Wood 2010). However, seemingly holistic conceptualisations suffer from several shortcomings that should not be ignored (Sarewitz 2010). As "the only truly holistic representation of a system would be the system itself," (p. 69) any representation is bound to be incomplete, and holistic frameworks can only capture sub-wholes, in which there is still room for disagreement and new fragmentation. For similar reasons, wholeness in technological apparatuses is impossible too. Even seemingly unified infrastructures such as the Internet rely physically on a heterogeneous patchwork of technologies, which Dourish and Bell (2011) simply call a "mess." The technologies for the Digital Earth are indeed no exception, and are marked by a proliferation of incompatible operating systems, data formats, and software tools.

Although visions of social, epistemological, ecological, and spiritual wholeness cannot be fulfilled, they can set in motion valuable techno-scientific work. In limited contexts, indeed, new media actually enable a higher degree of wholeness, and the Digital Earth might facilitate interdisciplinarity and cohesion in Earth-related disciplines. However, new media, clashing with the complexities of the societies that generate and adopt them, can unexpectedly unleash social, epistemological, and psychological fragmentation, and the Digital Earth has no special status to be exempted from this phenomenon. Furthermore, increased media-supported knowledge about the Earth does not necessarily entail more control over "complex, interdependent, non-deterministic socio-technical systems that no one knows how to alter in particular ways to yield particular outcomes" needed to solve real social and environmental problems (Sarewitz 2010, p. 69).

While Carey and Quirk (1989a) called for the "demythologising" the rhetoric of these discourses, it can be argued that such efforts are futile. In most cases, utopian media futures promptly defuse themselves as the new technologies disappear into daily banality, replacing the sublime with the mundane and the familiar. Animating myths about media that are becoming part of daily life becomes increasingly difficult and, because of a well-known paradox of media history, the social impact of technologies is greatest once they mature out of the mythical phase, deeply altering society as banal, mostly invisible tools (Mosco 2004).

However, this is not to say that media futures cannot be harmful. In this sense, envisionments of new media futures are not dangerous because they might become true, as dystopians think, but because of the underlying discourses that they silently carry. Depoliticisation is the most severe risk of technological myths. Discourses of media wholeness can be deployed to cover what Vaidhyanathan (2011) terms "infrastructural imperialism," i.e. the imposition of particular media infrastructures across the globe outside democratic discussions, as in the case of Google's geo-technologies. As often happened in the context of political imperialism, grand narratives of universal technological progress can mask and justify less appealing local realities of domination. Furthermore, discourses of political wholeness can always degenerate into the erasure of fragmentation to enforce fictional totalities, as experienced in 20th century totalitarianisms.

As Carey and Quirk (1989a, pp. 105-6) noted, techno-utopian visions betray deep frustration with primarily non-technological issues to which no effective solution has been found. In this sense, these visions can divert tensions that should be more fruitfully articulated in the political arena. In this sense, the presentation of specific information technologies as natural and inevitable rather than politically charged can stave off democratic discussions about their adoption. Notably, privacy issues represent a crucial and unresolved node for the democratic regulation of Digital Earth technologies.

## 6. Conclusion

Fragments of the future are constantly generated everywhere. Sociologist Barbara Adam (2009) points out that futures "are produced by the breadth of social institutions: politics, law and the economy, science, medicine and technology, education, and religion … at the level of the individual, the family, social groups, companies, and nations" (p. 429). For media and mobility studies, visions of futures fostered by new technologies should be taken seriously, as they reveal deep assumptions, expectations, and beliefs about the development and adoption of new media. This article has analysed the Digital Earth vision, outlined by Al Gore in 1998 as a high-resolution digital model of the planet to enable scientific, political, educational, and recreational applications. The Digital Earth was then contextualised in a constellation of imagined futures in which new media would refine the relationship between humans and the planet. The deep core of such envisionments of proximate futures is the achievement of forms of *wholeness* in the face of perceived fragmentation.

Envisionments of proximate futures such as the Digital Earth fulfil several purposes. First, they establish simple and striking metaphors in techno-scientific landscapes dominated by complexities and disagreements, evoking awe for the technological sublime that proliferates in marketing discourses. These visions inspire both research agendas and a rhetorical tone that can be used by scientists and technologists to frame and give meaning to their work, staving off the suspicion of futility. Second, at the economic level, they play a "regulative" role, aligning interests, capital, and resources, promoting and steering the actual development and adoption of media technologies. Third, they can implicitly support broader political agendas, becoming metaphors of far-reaching transformations of society, hinging on discourses of inevitable progress and hopes for a brighter future.

As this article has shown, the vision of the Digital Earth can be interpreted as a techno-scientific myth. While Al Gore was the first bricoleur who animated the myth during his second term as U.S. Vice President, the myth would have died quickly if it was not fed. As geo-technologies become embedded in everyday life and disappear into the mundane, the Digital Earth needs to be constantly revived and updated by a community of scientists. As the myth promises to lead toward wholeness through a new medium, inevitably new fragmentation arises as a result of media adoption. Therefore, the quest for wholeness, such as David Bohm's, is bound to be ineffectual in the kaleidoscope of media landscapes, which thrive on shifting patterns of temporary equilibria and deep fragmentation. The Digital Earth vision, rather than solving deep contradictions in the relationship between human beings and the planet, acts as a reminder of the irreducible multiplicity of media futures, each bringing its own promise of wholeness and its real, inevitable chaos.

---

[1] The social and epistemic impact of geographic information systems (GIS) in the 1990s has been dissected in the context of "Critical GIS."

[2] Mosco (2004) sarcastically summarised these exaggerated visions in Gore's Law: "myths about the Internet double in their distance from reality every 18 months."

[3] http://www.digitalearth-isde.org